\documentstyle[12pt]{article}
\renewcommand{\thefootnote}{\fnsymbol{footnote}}
\newcommand{\bea}{\begin{eqnarray}}
\newcommand{\ena}{\end{eqnarray}}
\newcommand{\vs}[1]{\vspace{#1 mm}}

\newcommand{\z}{\omega}

\newcommand{\PL}[1]{Phys.\ Lett.\ {\bf #1}}

\newcommand{\PR}[1]{Phys.\ Rev.\ {\bf #1}}
\newcommand{\PRL}[1]{Phys.\ Rev.\ Lett.\ {\bf #1}}

\newcommand{\PTPS}[1]{Prog.\ Theor.\ Phys.\ Suppl.\ {\bf #1}}

\newcommand{\EPJ}[1]{Eur.\ Phys.\ J.\ {\bf #1}}

\begin{document}
\noindent
\topmargin 0pt
\oddsidemargin 5mm

\begin{titlepage}
\setcounter{page}{0}
\begin{flushright}
May,  2000\\
OU-HET 346\\
hep-ph/0005179\\
\end{flushright}
\vs{2}
\begin{center}
{\Large{\bf Democratic-type neutrino mass matrix}}
\footnote{Talk given by E. Takasugi at 
Workshop on Neutrino Oscillations and their Origin, 
Fuji-Yoshida, Japan, 11-13 Feb 2000}\\
\vs{6}
{\large
Takahiro Miura\footnote{e-mail address:
miura@het.phys.sci.osaka-u.ac.jp}
and Eiichi Takasugi\footnote{e-mail address:
takasugi@het.phys.sci.osaka-u.ac.jp}\\
\vs{2}
{\em Department of Physics,
Osaka University \\ Toyonaka, Osaka 560-0043, Japan} \\
\vs{6}
Masaki Yoshimura\footnote{e-mail address:
myv20012@se.ritsumei.ac.jp}\\
\vs{2}
{\em Department of Physics,
Ritsumeikan University \\ Kusatsu, Shiga 525-8577, Japan} }
\end{center}
\vs{5}
\centerline{{\bf Abstract}}  
We consider the democratic-type neutrino mass matrix and 
show that this matrix  predicts 
the atmospheric neutrino mixing to be almost maximal, 
$\sin^2 2\theta_{atm}>0.999$ as well as the large CP violation 
(the CP violation phase in the standard form is maximal 
$\delta=\pi/2$). We construct the $Z_3$ symmetric dimension 
five effective Lagrangian with two up-type Higgs doublets and 
show that this Lagrangian leads to  the democratic neutrino mass 
matrix. Furthermore, we consider the restricted model with 
one up-type Higgs doublet and obtain the prediction, 
$0.87<\sin^2 2\theta_{sol}<8/9$. 

\end{titlepage}
\newpage
\vskip 2cm
\renewcommand{\thefootnote}{\arabic{footnote}}
\setcounter{footnote}{0}

\section{Introduction}

The observation of the atmospheric neutrino by SuperKamiokande[1] 
has shown the existence of the neutrino masses and the 
neutrino mixings. In particular, the data show that the 
mixing between $\nu_\mu$ and $\nu_\tau$ is favored and 
$\sin^2 2\theta_{atm}\simeq 1$ and $ \Delta_{atm}^2 
\sim 3.5 \times 10^{-3}{\rm eV}^2$. 
The solar neutrino problem is now considered to be due to 
the $\nu_e$ and $\nu_{\mu}$ oscillation, but the information 
on masses and mixing angles are ambiguous. Now four 
solutions are given[2]. The another 
crucial information is given by CHOOZ group[3] that 
$|V_{13}|<0.16$

We interpret these in the neutrino mixing 
matrix in the standard form[4]
\bea 
V_{SF}= \pmatrix{c_{12}c_{13}&s_{12}c_{13}&s_{13}e^{-i\delta}\cr
 -s_{12}c_{23}-c_{12}s_{23}s_{13}e^{i\delta}&
  c_{12}c_{23}-s_{12}s_{23}s_{13}e^{i\delta}& s_{23}c_{13}
  \cr
  s_{12}s_{23}-c_{12}c_{23}s_{13}e^{i\delta}&
  -c_{12}s_{23}-s_{12}c_{23}s_{13}e^{i\delta}& c_{23}c_{13}
  \cr}\;,
\ena
where $s_{ij}=\sin \theta_{ij}$ and $c_{ij}=\cos \theta_{ij}$. 

The large atmospheric neutrino mixing  requires 
$|s_{23}|\simeq |c_{23}|\simeq 1/\sqrt{2}$
and the CHOOZ data gives the bound $|s_{13}|<0.16$. 

Now we face the following questions:
(1) Why $|s_{23}|\sim |c_{23}|\sim 1/\sqrt2$ ? 
(2) What is the size of $s_{12}$ ?
(3) Why $s_{13}$ is so small ? 
(4) What is the size of the CP violation phase $\delta$ ? 

It is quite hard to construct the neutrino mass matrix which 
answers all these questions. In this note, we focus on 
the questions, 
(1), (2) and (4), by using the democratic-type mass matrix 
for the neutrino[5],[6].  Throughout of this note, we consider 
the neutrino mass matrix in the basis where charged leptons 
mass matrix is diagonal.  

\section{The model construction}

We are interested in  the CP violation phase. 
A famous model that predicts the CP violation phase is 
the Tri-maximal mixing scheme[7]  
\bea
V_T=\frac{1}{\sqrt{3}}\pmatrix{
1 & 1 & 1 \cr
\z & \z^2 & 1 \cr
\z^2 & \z & 1
}\;, 
\ena
where $\z$ is the element of $Z_3$ and is given by 
$\z=e^{i2\pi/3}$, i.e., $\z^3=1$. This matrix 
 predicts $|V_{13}|=1/\sqrt{3}$ in  conflict  
with the CHOOZ data[3]. However, this scheme has an 
quite interesting predictions; (i) The maximal CP phase, 
$\delta=\pi/2$ and (ii) 
the maximal CP violation,  
$|J_{CP}|_{Tri}=1/6\sqrt{3}$. 

\vskip 1mm
\noindent
(a) Deformation from the Tri-maximal mixing 

We considered the deformation of the Tri-maximal mixing 
matrix by the orthogonal matrix $O$, $V=V_TO$ and 
 found  quite significant predictions[5],  
\bea
|s_{23}|=|c_{23}|=\frac1{\sqrt 2} \;,
\qquad \delta=\frac{\pi}2\;.
\ena 

\noindent
(b) The possible origin of the mixing matrix in the form of 
$V=V_TO$
 
We define the new basis ($\psi$ basis) which is related to 
the flavor eigenstate basis by 
$(\psi_1,\psi_2,\psi_3)^T=V_T^\dagger(\nu_{e L},\nu_{\mu L},
\nu_{\tau L})^T$. Now, the matrix to realize 
the mixing matrix $V=V_TO$ is some real matrix  
in the $\psi$ basis, 
\bea  
 V_T^T m_\nu V_T= \pmatrix{
   m^0_1+\tilde{m}_1&\tilde m_3&\tilde m_2\cr
   \tilde m_3&m^0_2+\tilde{m}_2&\tilde m_1\cr
   \tilde m_2&\tilde m_1&m^0_3+\tilde{m}_3\cr}\;,
\ena
where $m_i^0$ and $\tilde m_i$ are real parameters. 
By inverting, we obtain $m_\nu$ 
in the flavor eigenstate basis as
\bea
m_\nu &=&
\frac{m^0_1}3\pmatrix{1&\z^2&\z\cr \z^2&\z&1\cr 
         \z&1&\z^2}+
\frac{m^0_2}3\pmatrix{1&\z&\z^2\cr \z&\z^2&1\cr 
         \z^2&1&\z\cr }+
\frac{m^0_3}3 \pmatrix{1&1&1\cr 1&1&1\cr 1&1&1\cr}
\nonumber\\
&&+
\tilde{m}_1\pmatrix{1&0&0\cr 0&\z&0\cr 0&0&\z^2\cr}+
\tilde{m}_2\pmatrix{1&0&0\cr 0&\z^2&0\cr 0&0&\z\cr}+
\tilde{m}_3\pmatrix{1&0&0\cr 0&1&0\cr 0&0&1\cr}\;,
\ena
which we called the democratic-type mass matrix[5]. 

\vskip 1mm
\noindent
(c) The dimension five effective Lagrangian with the 
$Z_3$ symmetry

The democratic-type mass matrix is derived from the Lagranfian 
by imposing the $Z_3$ symmetry. 
We define the irreducible representations of $Z_3$ as 
$\Psi_1=\frac1{\sqrt 3}(\ell_e+\z^2 \ell_\mu+\z \ell_\tau)$, 
$\Psi_2=\frac1{\sqrt 3}(\ell_e+\z \ell_\mu+\z^2 \ell_\tau)$ and 
$\Psi_3=\frac1{\sqrt 3}(\ell_e+\ell_\mu+\ell_\tau)$, 
where $\ell_i$ is the left-handed lepton doublet defined by, 
say,  $\ell_e^T=(\nu_{eL}, e_L)$. 
The fields $\Psi_i$ transform  under 
the permutation  $\ell_e$, $\ell_\mu$ and $\ell_\tau$, 
$\Psi_1\to \z \Psi_1$, $\Psi_2\to \z^2 \Psi_2$ and 
$\Psi_3\to \Psi_3$. 
With the definition, 
$\Psi_i=(\psi_i, e_i)$, the above relation is viewed 
as the transformation from the flavor eigenstate basis to 
the $\psi$ basis. 

If we introduce two up-type Higgs doublets that behave as 
$H_{u1} \rightarrow \z^2 H_{u1}$ and  $H_{u2} \rightarrow \z H_{u2}$, 
then we can construct the $Z_3$ invariant dimension five 
effective Lagrangian as 
\bea
{\cal L}_{\rm y}
&=&-
  \Biggl(
     (m^0_1+\tilde{m}_1)\overline{(\Psi_1)^C} \Psi_1 
                        \frac{H_{u1} H_{u1}}{u^2_1}+
     (m^0_2+\tilde{m}_2)\overline{(\Psi_2)^C} \Psi_2 
                        \frac{H_{u2} H_{u2}}{u^2_2}  \nonumber\\
     &&\mbox{ }
      +(m^0_3+\tilde{m}_3)\overline{(\Psi_3)^C} \Psi_3 
                        \frac{H_{u1} H_{u2}}{u_1u_2}
\Biggr)\nonumber\\
&&-2\left(
     \tilde{m}_1\overline{(\Psi_2)^C} \Psi_3 \frac{H_{u1}
H_{u1}}{u^2_1}+
     \tilde{m}_2\overline{(\Psi_1)^C} \Psi_3 \frac{H_{u2}
H_{u2}}{u^2_2}+
     \tilde{m}_3\overline{(\Psi_1)^C} \Psi_2 \frac{H_{u1}
H_{u2}}{u_1u_2}  
  \right)\;,
\ena
where $u_i$ is the vacuum expectation value of the neutral 
component of $H_{ui}$. 

After the Higgs fields acquire the vacuum expectation values, 
the neutrino mass matrix in the $\psi$-basis defined by Eq.(4), 
and thus the democratic-type neutrino mass matrix in Eq.(5) 
is obtained.  

\section{A restricted model -one Higgs case-}

Here we consider the case with only one up-type Higgs by  
keeping $H_{u1}$. Then, from Eq.(6), we only have the 
$\tilde m_1$ and $m_1^0$ terms.  
As we can see from Eq.(4) with $\tilde m_2=\tilde
m_2=0$ and $m_2^0=m_3^0=0$, 
we observe that the matrix is diagonalized by 
\bea
V_1&=&V_T\pmatrix{1&0&0\cr0&\frac1{\sqrt 2}&-\frac1{\sqrt 2}\cr
           0&\frac1{\sqrt 2} & \frac1{\sqrt 2}\cr}
    =\pmatrix{1&0&0\cr0&\z&0\cr 0&0&\z^2}
 \pmatrix{\sqrt{\frac13}&-\sqrt{\frac23}&0\cr 
            \frac{1}{\sqrt 3}&\frac{1}{\sqrt 6}&-\frac{1}{\sqrt 2}\cr
            \frac{1}{\sqrt 3}&\frac{1}{\sqrt 6}&\frac{1}{\sqrt 2}\cr}
   \pmatrix{1&0&0\cr 0&-1&0\cr 0&0&i\cr}\;.
\nonumber           
\ena  
This mixing matrix is a new type and predicts 
$\sin^2 2\theta_{sol}=\frac89$ and $\sin^2 2\theta_{atm}=1$
in contrast to  the Bi-maximal mixing[8] which predicts 
$\sin^2 2\theta_{sol}=1$ and 
$\sin^2 2\theta_{atm}=1$, and the democratic mixing[9] does 
$\sin^2 2\theta_{sol}=1$ and $\sin^2 2\theta_{atm}=8/9$. 

Unfortunately, this model predicts the degenerate masses 
$m_2=-m_3(=\tilde m_1)$. In order to remedy this deficit, 
we introduce the symmetry breaking terms which preserve 
the $Z_2$ symmetry 
$\Psi_1 \to - \Psi_1$. 
Among the interaction terms in Eq.(14), the $Z_2$ symmetry 
excludes $\tilde m_2$ and $\tilde m_3$ terms and thus we obtain 
the neutrino mass matrix including four parameters $\tilde m_1$, 
$m_1^0$,  $m_2^0$ and $m_3^0$. This matrix is diagonalized by[6] 
\bea
V= \pmatrix{1&0&0\cr0&\z&0\cr 0&0&\z^2}
   \pmatrix{\frac1{\sqrt{3}}&-\sqrt{\frac23}c'&i\sqrt{\frac23}s'\cr 
            \frac 1{\sqrt 3}&\frac1{\sqrt 6}(c'+i\sqrt 3 s')
            &-\frac1{\sqrt 6}(\sqrt 3 c'+is')\cr
            \frac1{\sqrt 3}&\frac1{\sqrt 6}(c'-i\sqrt 3 s')
            &-\frac1{\sqrt 6}(\sqrt 3 c'-is')\cr}
   \pmatrix{1&0&0\cr 0&-1&0\cr 0&0&i\cr}\;,
\ena
where $s'=\sin \theta'$, $c'=\cos \theta'$ and 
$\tan \theta'=\Delta_-/(
\tilde m_1+\sqrt{\tilde m_1^2+ \Delta_-^2})$. 
The neutrino masses are given by 
$m_1=m_1^0+\tilde m_1$, $
m_2=m_2^0+\Delta_- +\sqrt{\tilde m_1^2+ \Delta_-^2}$, 
$m_3=m_2^0+\Delta_- -\sqrt{\tilde m_1^2+ \Delta_-^2}$. 
 
From the mixing matrix in Eq.(7), we find 
$\delta=\frac{\pi}2$ and 
$\tan \theta_{12}=\sqrt{2-3\sin^2 \theta_{13}}$, 
while $s_{13}$ is left 
a free parameter. Now we have 
$\sin^2 2\theta_{sol}= \frac 89 c'^2$ and $
\sin^2 2\theta_{atm}= 1-\frac 49 s'^4$.
If we impose the CHOOZ bound, $|\sqrt{2/3}s'|<0.16$, we have 
$0.87<\sin^2 2\theta_{sol}<8/9$ and 
$\sin^2 2\theta_{atm}>0.999$. 
If the future precision experiments on the neutrino mixing 
show that the atmospheric neutrino mixing is very close to 
the maximal value and the solar neutrino mixing is large but not 
maximal, our model would be a very good candidate. 

Here, the phase matrix ${\rm diag}(1,\z,\z^2)$ is absobed into 
charged lepton phases and the phase matrix ${\rm diag}(1,-1,i)$ 
represents  the CP violating Majorana phase matrix[10],[11], which 
is relevant to the purely lepton number violationg processes such 
as the neutrinoless double beta decay[12]. 

As for the CP violation, the Jarlskog parameter is
\bea 
\frac{|J_{CP}|_{our\; model}}{|J_{CP}|_{max}} =
\sqrt{ 6}
\left | s_{13}c_{13}^2\sqrt{1-\frac 32 s_{13}^2}\right |<0.37\;,
\ena
where $|s_{13}|=\sqrt{2/3}s'$. The reduction rate from the 
maximum CP violation is solely dependent on the mixing angle 
$s_{13}$. If we use the CHOOZ data $|s_{13}|<0.16$, we obtain 
the ratio is smaller than 0.37, which is good enough to be 
observed in the  near future laboratory  experiments on the 
neutrino oscillations.

\end{document}